\documentclass[11pt]{article}
\usepackage{geometry}          
\usepackage[labelfont=bf]{caption} 
\usepackage[font={small}]{caption}   

\usepackage{graphicx}  
\usepackage{dcolumn}   
\usepackage{bm}        
\usepackage{amssymb}   
\usepackage{adjustbox}

\usepackage{amsmath}   
\usepackage{paralist}  

\usepackage[labeled]{multibib}
\newcites{S}{Supplementary References}

\usepackage[labelfont=bf]{caption} 
\usepackage[strict]{changepage} 
\usepackage{textgreek} 

\newcommand{\am}{\ensuremath{a_{-}} }

\newcommand{\gap}{\ensuremath{\gamma_{a+}} }
\newcommand{\gam}{\ensuremath{\gamma_{a-}} }

\usepackage[usenames,dvipsnames]{color}    
\usepackage{soul}   
\setstcolor{red} 
\setul{}{1.5pt}  

\def\be{\begin{equation}}      
\def\ee{\end{equation}}
\def\bea{\begin{eqnarray}}      
\def\eea{\end{eqnarray}}
\def\beau{\begin{eqnarray*}}      
	\def\eeau{\end{eqnarray*}}
\usepackage{amsmath,amsfonts,amssymb}
\usepackage{wrapfig}
\usepackage{graphicx}
\usepackage{bbm}
\usepackage{graphics}

\def \beq {\begin{equation}}
\def \eeq {\end{equation}}
\def \ba {\begin{eqnarray}}
\def \ea {\end{eqnarray}}

\def\beu{\begin{equation*}}   
\def\eeu{\end{equation*}}

\providecommand{\mr}[1]{\mathrm{#1}}
\providecommand{\abs}[1]{\left\lvert#1\right\rvert}   

\newcommand{\beginsupplement}{%
	\setcounter{table}{0}
	\renewcommand{\thetable}{S\arabic{table}}%
	\setcounter{figure}{0}
	\renewcommand{\thefigure}{S\arabic{figure}}%
	\setcounter{page}{1}
	\renewcommand{\thepage}{S\arabic{page}}%
	\setcounter{section}{0}
	\renewcommand{\thesection}{S\arabic{section}}%
	\setcounter{equation}{0}
	\renewcommand{\theequation}{S\arabic{equation}}%
}

\usepackage{lipsum}
\usepackage{placeins}

\usepackage{adjustbox}
\usepackage{titling}

\begin{document}

	\title{Dynamically induced robust phonon transport and chiral cooling in an optomechanical system}
	
	\author{Seunghwi Kim$^1$, Xunnong Xu$^2$, \\ Jacob M. Taylor$^{2,3\ast}$, and Gaurav Bahl$^{1\ast}$\\
		\\
		\footnotesize{$^1$ Mechanical Science and Engineering, University of Illinois at Urbana-Champaign}\\
		\footnotesize{Urbana, Illinois 61801, USA}\\
		\footnotesize{$^2$ Joint Quantum Institute, University of Maryland, }\\
		\footnotesize{College Park, Maryland 20742, USA}\\
		\footnotesize{$^3$ Joint Center for Quantum Information and Computer Science,} \\
		\footnotesize{National Institute of Standards and Technology, Gaithersburg, Maryland 20899, USA}\\
		\footnotesize{$^\ast$ To whom correspondence should be addressed; bahl@illinois.edu, jmtaylor@umd.edu} \\
	}
	
	\date{}
	
	\vspace*{-2cm}
	{\let\newpage\relax\maketitle}

	\begin{abstract}

		The transport of sound and heat, in the form of phonons, can be limited by disorder-induced scattering. In electronic and optical settings the introduction of chiral transport, in which carrier propagation exhibits parity asymmetry, can remove elastic backscattering and provides robustness against disorder. However, suppression of disorder-induced scattering has never been demonstrated in non-topological phononic systems. Here we experimentally demonstrate a path for achieving robust phonon transport in the presence of material disorder, by explicitly inducing chirality through parity-selective optomechanical coupling. We show that asymmetric optical pumping of a symmetric resonator enables a dramatic chiral cooling of clockwise and counterclockwise phonons, while simultaneously suppressing the hidden action of disorder. Surprisingly, this passive mechanism is also accompanied by a chiral reduction in heat load leading to optical cooling of the mechanics without added damping, an effect that has no optical analogue. This technique can potentially improve upon the fundamental thermal limits of resonant mechanical sensors, which cannot be attained through sideband cooling.

	\end{abstract}

	\newpage

	\section{Introduction}

	Efforts to harness the optical and mechanical properties of achiral resonators are leading to new approaches for quantum noise limited sources \cite{Kippenberg555,Li309,Loh:15}, preparation of quantum states of matter \cite{Gigan:2006p1091,Arcizet:2006hv,Chan_GroundState_2011,Verhagen:2012ei}, and ultra-high precision metrology \cite{Teufel:2009gn,PhysRevA.82.061804,Krause:2012cf,Gavartin:2012el}.
	Since all these efforts are aided by long coherence times for resonant excitations, they are fundamentally limited by structural disorder, even in systems with high symmetry, and by thermal noise  in the mechanics.
	While optomechanical sideband cooling can lower the effective temperature of a mechanical oscillator, it does not modify the heat load, and thus does not fundamentally modify the contribution of thermal noise for, e.g., sensing or transduction \cite{Krause:2012cf,Gavartin:2012el}.
	Surprisingly, the chiral edge states of topological insulators -- in which different parity excitations travel in different directions -- can provide improved transport properties, giving rise to unique physics ranging from nonreciprocal wave propagation to disorder-free transport in quantum Hall systems \cite{Halperin:1982tb,Wang09,Hafezi:2013jg,Susstrunk15,Peano:2016je}.
	At the same time, inducing nonreciprocal behavior by breaking parity-symmetry in achiral non-topological devices forms the basis for circulators \cite{Fleury:2014fn,Estep:2014jp} and recent proposals for optomechanical isolation \cite{Hafezi:2013jg,Kim2015,Dong2015,Peano:2016je}.
	However, experiments to date on nonreciprocal optomechanical devices \cite{Kim2015, Dong2015, Shen:gt} have focused entirely on optical behavior, and there has been no direct exploration on the chiral nature of propagating phonons in such systems, nor on the disorder tolerance induced through chirality.

	Here we show how to optically impart chirality to achiral mechanical systems.
	Our approach results in disorder-less transport of sound, simultaneously improving the isolation of phonon modes from their bath and lowering their heat load without added damping \cite{XuTaylor_2016}.
	We use a particularly simple class of systems for examining chiral behavior: passive devices with degenerate 
	forward- and backward-propagating modes, such as ring cavities and whispering-gallery resonators (WGRs).
	The modes of these structures can alternatively be described as opposite parity pairs having clockwise ($+$ or cw) and counter-clockwise ($-$ or ccw) pseudo-spin of circulation \cite{Hafezi:2013jg}.
	Chiral here indicates pseudo-spin-dependent behavior in the system.
	Disorder breaks parity conservation in WGRs, preventing chiral behavior and leading to the additional loss of energy in high-$Q$ modes via pseudo-spin flips and scattering into bulk modes, both optically \cite{Gorodetsky:2000vr} and mechanically \cite{Knopoff:1964ve, Pao:1976wc}.
	Recent work has shown that asymmetric optical pumping of one pseudo-spin direction in WGRs explicitly introduces chiral behavior for photons with the assistance of even weak optomechanical coupling \cite{Kim2015}.
	In this work, we demonstrate that this induced symmetry breaking
	in fact imparts parity-dependent behavior throughout the system, the chiral echoes of which are observable across its mechanical properties as the system develops robustness to parity-breaking disorder.

	We observe three significant phononic chiral effects.
	First, cw and ccw phonons experience dramatic chiral optomechanical cooling. While this may be anticipated from past experiments on Brillouin optomechanical coupling with traveling phonons \cite{Bahl:2012jm, Kim2015}, such chiral phonon propagation has never been experimentally reported.
	Second, the optomechanical damping selectively provided to cw phonons results in 
	mitigation of the disorder-induced loss for phonons having the opposite (ccw) pseudo-spin -- effectively a phononic analogue of the quantum Zeno effect. 
	Finally, while the cw phonon modes experience conventional optomechanical cooling \cite{Gigan:2006p1091,Arcizet:2006hv,marquardt.2008,park.2009,Chan_GroundState_2011,Bahl:2012jm}, an isolated high-$Q$ ccw mode simultaneously experiences a reduction in damping and a reduction in temperature.
	This result reveals a surprising form of optomechanical cooling that occurs through chiral refrigeration of the thermal bath composed of the (cw) bulk mechanical modes of the system.

	\vspace{12pt}

	\begin{figure}[p!]
		\begin{adjustwidth}{-1in}{-1in}
			\includegraphics[width=0.65\textwidth]{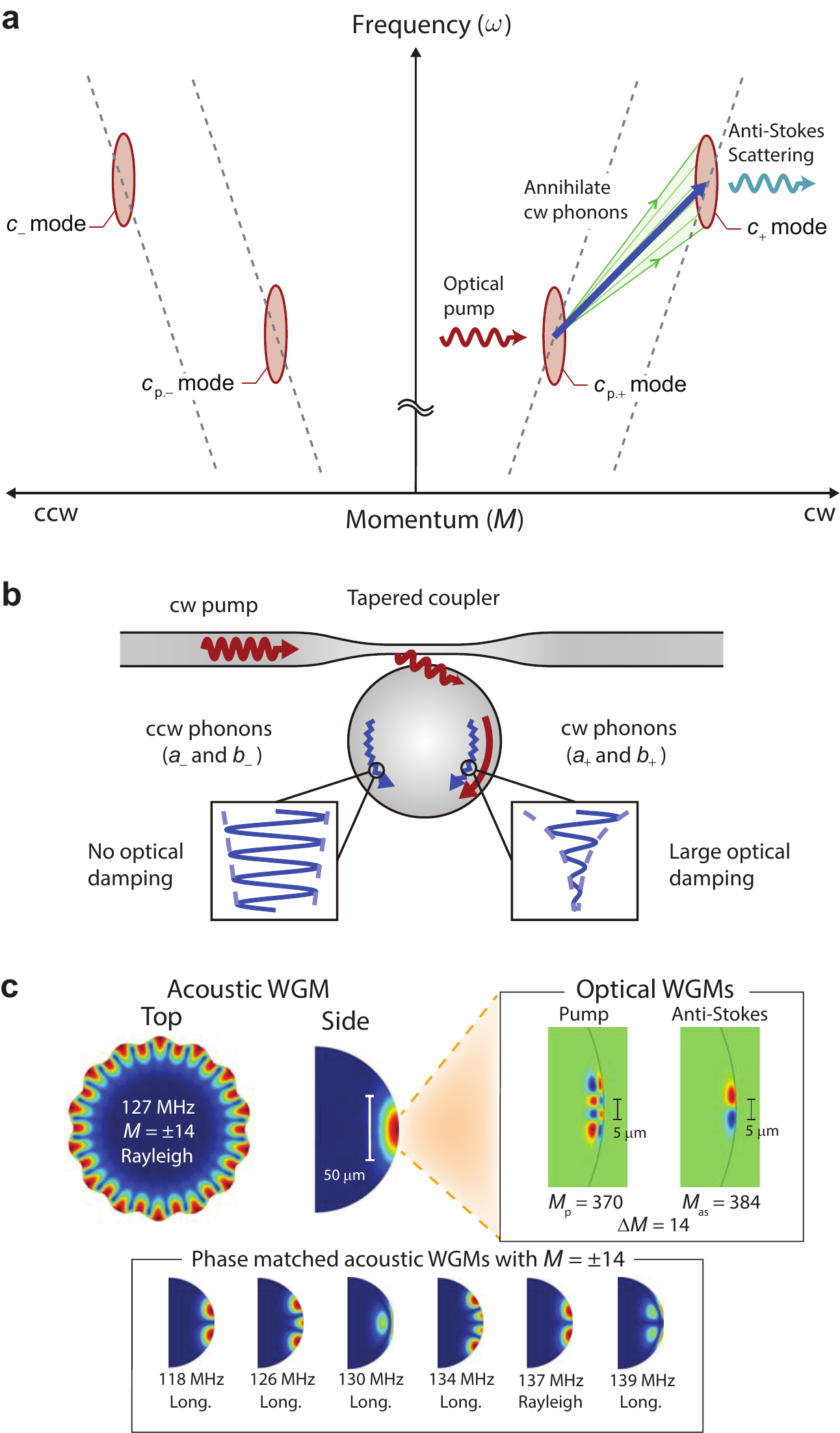}
			\centering
			\caption{
				\textbf{Chiral cooling and modal relationships in a whispering-gallery resonator.} 
				\textbf{(a)} Configuration of the two requisite optical whispering-gallery modes (WGMs) with cw ($+$) and ccw ($-$) degeneracy is illustrated in ($\omega, M$) space. Integer $M$ describes the azimuthal momentum of WGMs. Anti-Stokes Brillouin scattering from cw pumping of the lower mode annihilates only cw phonons of all phase-matched dispersion relations, while ccw phonons remain nominally unaffected, thus breaking parity symmetry in the system. The optical modes are illustrated elongated in $\omega$-space due to the extremely large phase velocity of photons in comparison to phonons. 
				\textbf{(b)} Directional optical interface to the resonator modes is achieved via tapered optical fiber. Unidirectional optical pumping results in dramatic chiral damping of the phonons. $a_{\pm}$ and $b_{\pm}$ phonon modes are described in Fig.~\ref{fig:model}.
				\textbf{(c)} Our experiment is performed using a high-$Q$ $M=\pm 14$ acoustic WGM ($a_{\pm}$) using two optical WGMs having the same momentum separation. There also exist an extended group of acoustic bulk modes having the same $M$ (representative members illustrated) that are also phase-matched but are not observable above the noise floor due to high mechanical damping and poorer frequency-matching. These are considered in our model collectively as a quasi-mode ($b_{\pm}$).
			}
			\label{fig:1}
		\end{adjustwidth}
	\end{figure}
	
	\section{Results}
	\subsection{Modal relationships in a whispering-gallery resonator}
	The cylindrical symmetry of our system means that the whispering-gallery modes (WGMs)  take the functional form $f(r,z) e^{i(M\phi - \omega t)}$, 
	where $\omega$ is the eigenfrequency and signed integer $M$ describes the propagation momentum or azimuthal order on the angular spatial variable $\phi$. The transverse mode profile is given by $f(r,z)$.
	We now introduce the specific mechanical and optical WGMs that participate in our experiment, and the nature of their interaction.
	Our structure hosts frequency-adjacent optical modes belonging to different families that may be populated with photons of either pseudo-spin (Fig.~\ref{fig:1}a), for a total of four optical modes in the experiment. 
	Scattering between these optical modes is only permitted through acousto-optic coupling (Brillouin scattering) 
	via propagating phonons that match their frequency and momentum difference \cite{Bahl2011a}, which is termed the phase matching requirement. Thus, photon modes having cw (ccw) pseudo-spin can only be coupled through scattering from co-propagating phonons having cw (ccw) pseudo-spin.
	In this situation when we pump the lower-energy optical mode, anti-Stokes scattering to the higher mode annihilates phonons \cite{Bahl:2012jm} of matched parity, resulting in added optomechanical damping and cooling of the phonon mode. This system should thus exhibit significant optically-induced chirality in the transport of phonons that are phase matched for this interaction (Fig~\ref{fig:1}a,b). 
	A large family of mechanical WGMs satisfy these phase matching requirements, of which a few representative members are illustrated in Fig.~\ref{fig:1}c. The lowest transverse-order Rayleigh phonon mode is most likely to be observable as it interacts least with the supporting stem of the microsphere, i.e. features the highest $Q$-factor, and thus generates the strongest scattering between the optical modes \cite{Bahl2011a}. 
	However, disorder induced scattering can couple this high-$Q$ mode to the large population of lower-$Q$ bulk modes of the resonator, as they have the same azimuthal order $M$ but differ in extension into the bulk.
	Since the transverse optical mode profiles are much smaller than the transverse profiles of the mechanical modes (Fig~\ref{fig:1}c), the bare optical mode coupling to each of these many mechanical (physical) modes is of a similar order. 
	Below, we collectively treat the remaining lower-$Q$ bulk modes as a pair of phonon ``quasi-modes'' having large dissipation rate and fixed pseudo-spin. These quasi-modes act as a thermal bath for their parity-flipped high-$Q$ mode, while also directly coupling to the light.

	\subsection{Experimental demonstration of chiral cooling}
\begin{figure}[t!]
		\begin{adjustwidth}{-1in}{-1in}
			\includegraphics[width=\textwidth]{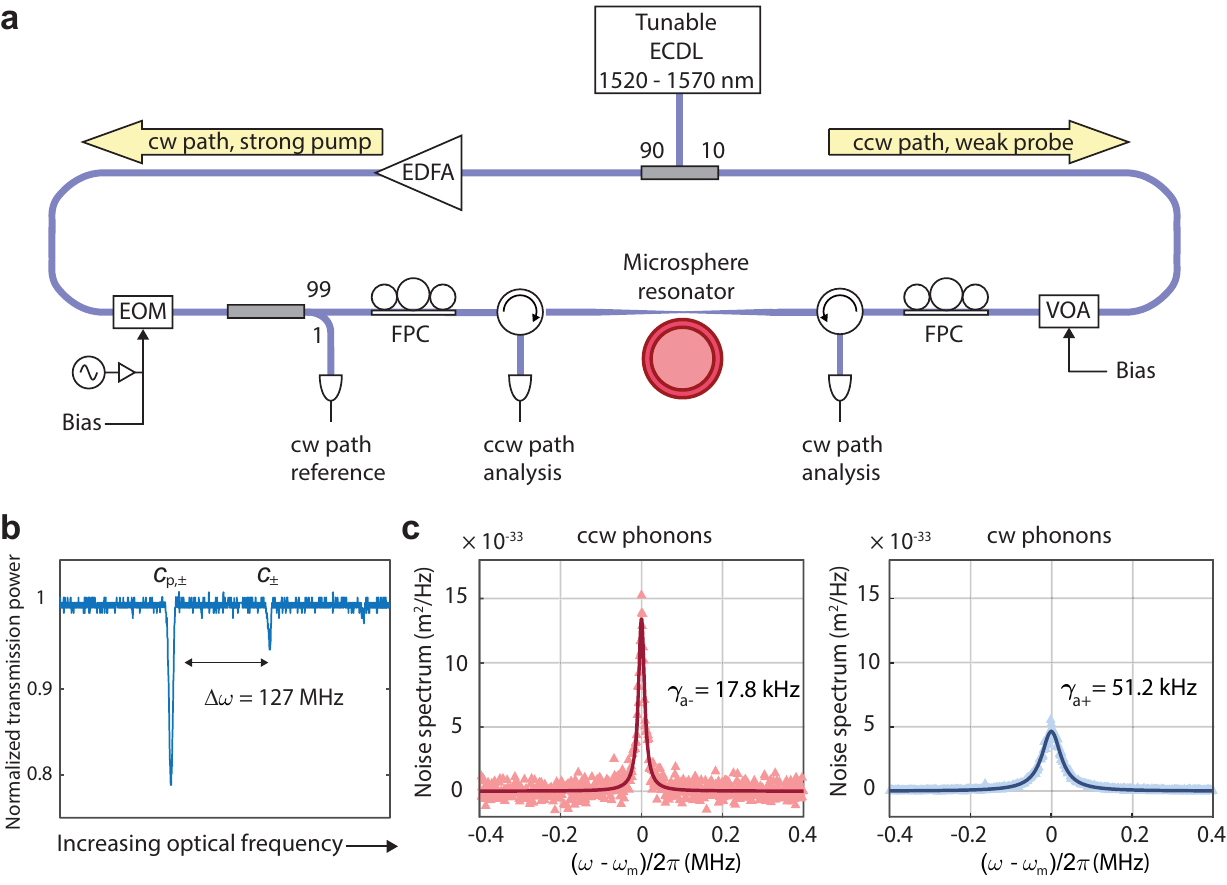}
			\centering
			\caption{
				\textbf{Measurement setup for chiral optomechanical refrigeration:} 
				\textbf{(a)} We perform the experiment using a silica whispering-gallery microsphere resonator that is interfaced via tapered optical fiber. A 1520 nm to 1570 nm tunable external cavity diode laser (ECDL) generates the asymmetric cw pump (strong) and ccw probe (weak) sources. An Erbium-doped fiber amplifier (EDFA) controls the cw pump power, for optomechanical cooling \cite{Bahl:2012jm} and for monitoring the cw phonon spectrum. An electro optic modulator (EOM) is employed to measure detuning of anti-Stokes scattered light from its optical mode through via induced transparency \cite{Kim2015}. A variable optical attenuator (VOA) is used to control the ccw probe. Fiber polarization controllers (FPC) are used to optimize coupling between the fiber and resonator in both directions.
				\textbf{(b)} The optical modes are mapped by measuring dropped power from the tapered fiber.
				\textbf{(c)} Dramatic chiral cooling of the high-$Q$ ($\omega_m = 2\pi \times 127$ MHz) phonon populations is immediately observed with strong cw pumping (140.9 $\mu \text{W}$). ccw phonons are also cooled slightly due to the ccw probe. Solid lines are Lorentzian fits to the data.
			}
			\label{fig:2}
		\end{adjustwidth}
	\end{figure}

	Our experiments are performed with a silica WGR of diameter $\sim$\,135\,$\mu \text{m}$ at room temperature and atmospheric pressure, using a tapered fiber coupler for optical interface at 1550 nm (Fig.~\ref{fig:2}a,b).
	The FWHM of both optical modes is approximately $\kappa = 5.1$ MHz, and the approximate simulated mode shapes are illustrated in Fig.~\ref{fig:1}c.
	Direct coupling between cw and ccw pseudo-spins, e.g. through optical Rayleigh scattering, is negligible for either optical mode in our experiment.
	The high-$Q$ mechanical mode is also a whispering-gallery mode at 127 MHz with azimuthal order of $M=14$ and mode shape illustrated in Fig.~\ref{fig:1}c. 
	Verification of the Brillouin phase-matching between these modes can be performed by means of forward Brillouin lasing \cite{Bahl2011a} and induced transparency measurements \cite{Kim2015}.
	To examine the potential for modification of the high-$Q$ phonon behavior, and the possibility of chiral transport of sound, we set up the experiment with two optical sources tuned to the lower optical mode in the clockwise (cw) and counter-clockwise (ccw) directions. 
	The role of the stronger cw ``pump'' is to induce cooling of the cw propagating phonons, while the role of the much weaker ccw ``reverse probe'' is to measure, via optical scattering,  the counter-propagating phonon behavior. The RF beat spectrum generated between the scattered light and the corresponding source in either direction provides a direct measure of the phonon mode spectrum (Supplementary Note 4), with sample measurements shown in Fig.~\ref{fig:2}c. In the experiment, the optical pump and probe sources are both derived from the same laser and are thus always at identical frequencies. Throughout the remainder of this work, no pump or probe field is delivered to the upper optical mode, in order to prevent coherent amplification of the phonons via Stokes Brillouin scattering.

	\vspace{12pt}

	Our first task is to measure the bare linewidth $\gamma_m$ of the high-$Q$ phonon mode without any optical pumping. We note that the bare linewidth ($\gamma_m$) of high-$Q$ mechanical WGMs may be qualitatively distributed into two forms of loss: those that maintain parity, which we call intrinsic dissipation ($\gamma$), and those that break parity, leading primarily to radiative damping via low-$Q$ bulk modes. 
	Thus $\gamma_m$ is traditionally the minimum measurable linewidth in any optomechanical sideband cooling experiment. We perform this measurement by detuning the source laser from the optical resonance such that little to no optomechanical cooling is induced by either the pump or the probe. For zero input power, the bare linewidth $\gamma_m = 12.5 (\pm 1.0)$ kHz is estimated by fit-based extrapolation of measured cw and ccw phonon linewidths $\gap$ and $\gam$ using the theoretical model (Supplementary Eqns.~5). We also obtain the single phonon optomechanical coupling strength $h_0 \approx 14 (\pm 2.5)$ Hz. All uncertainties in this manuscript correspond to 95\,\% confidence bounds of the fitted value.

	The optomechanical cooling rate can be controlled by the detuning $\mathit{\Delta}_2$ of the anti-Stokes scattered light from its optical mode, which we measure directly using the Brillouin Scattering Induced Transparency \cite{Kim2015}. In Fig.~\ref{fig:main}a we plot measurements of both cw and ccw phonon linewidth as a function of this detuning. We immediately see a striking direction-dependence of the damping rates of the cw and ccw phonons, that has never previously been reported. This chiral damping of phonons is a direct result of the momentum conservation rules that underly the Brillouin scattering interaction, and will not generally be available in traditional single-mode optomechanical systems. We note also that the relative power of the cw pump and ccw probe lasers is $\sim\,9:1$, so there is some sideband cooling of the \am phonons as well. 

	\vspace{12pt}

	\begin{figure}[t!]
		\begin{adjustwidth}{-1in}{-1in}
			\includegraphics[width=0.75\textwidth]{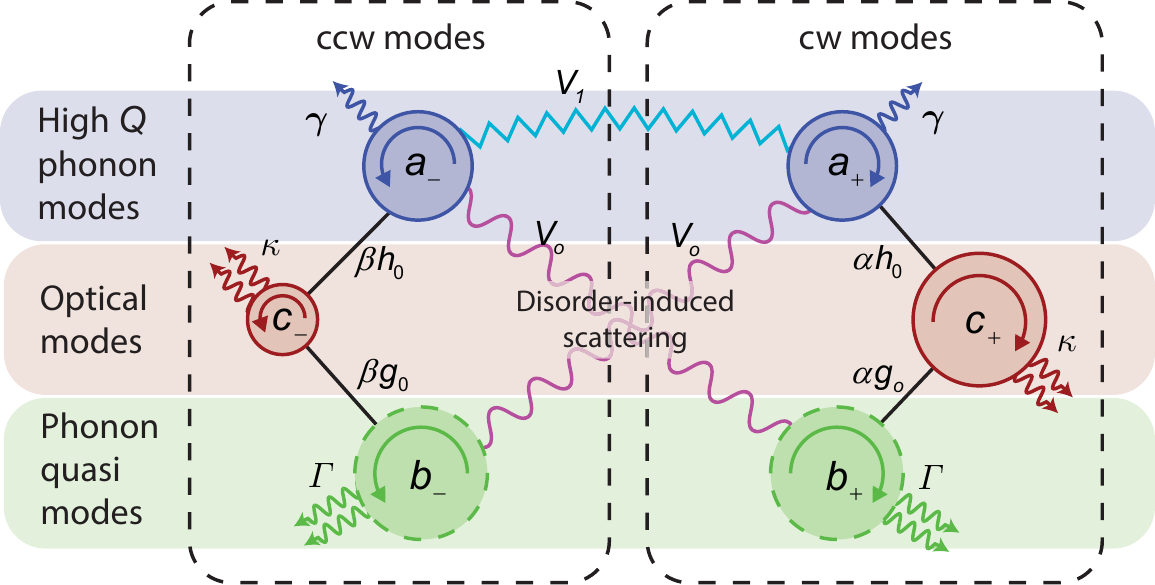}
			\centering
			\caption{
				Model for coupling between the anti-Stokes (higher frequency) optical modes $c_\pm$, high-$Q$ phonon modes $a_\pm$, and phonon quasi-modes $b_\pm$ in cw and ccw directions. Light couples to the $a_\pm$ and $b_\pm$ phonons with different optomechanical interaction strength. Disorder-induced scattering between $a_\pm$ phonons and the $b_\mp$ quasi-mode occurs with strength $V_0$, and between $a_\pm$ and $a_\mp$ high-$Q$ phonons occurs with strength $V_1$. Our experimental case features $V_1 \ll V_0$. Details on individual parameters are provided in the text.
			}
			\label{fig:model}
		\end{adjustwidth}
	\end{figure}

	\subsection{Model for chiral cooling}
	We now propose a model, detailed in Ref.~\cite{QuasiModeTheory_2016}, that incorporates all the essential physics described above and illustrated in Fig.~\ref{fig:model}. Specifically,  we define the higher frequency optical modes (anti-Stokes) through annihilation operators $c_\sigma$, with parity $\sigma=+$ for cw photons and $\sigma=-$ for ccw photons. These optical modes couple to the matched high-$Q$ phonon mode $a_\sigma$, via direct optomechanical interaction with strength $\alpha h_0$ (or $\beta h_0$), and also couple to the quasi-mode $b_\sigma$ with strength $\alpha g_0$ (or $\beta g_0$). In all cases these interactions conserve parity $\sigma$.
	$h_0$ and $g_0$ are the bare (single photon/phonon) optomechanical coupling strengths, while $\alpha$ and $\beta$ are the square root of the intracavity photon number in cw and ccw direction respectively due to the pump and probe lasers. We note that the optomechanical coupling to the quasi-mode (itself comprised of many low-$Q$ modes) can be large, due to the mode overlap highlighted in Fig.~\ref{fig:1}.

	To see the role chirality plays, we now also include the existence of disorder-induced scattering between $a_{\sigma}$ and $b_{\bar{\sigma}}$ modes having strength $V_0$ -- a term that explicitly breaks the conservation of parity. 
	We note that the definition of our modes already account for low-angle scattering that conserves parity but leads to damping. Thus we do not include a direct $a_\sigma$ to $b_\sigma$ term in our theory, as it is included in the definition of intrinsic linewidths $\gamma$ (for $a_\sigma$) and $\mathit{\Gamma}$ (for $b_\sigma$).
	We can thus represent this toy model system with the interaction Hamiltonian expression given by:
	\begin{align}
	H_{\text{int}} = & ~ \alpha c_+^{\dag} (h_0 a_{+}  + g_0 b_{+}) 
	+ \beta c_-^{\dag} (h_0  a_{-}  + g_0 b_{-})  \notag \\
	&+ V_0 (a_{-}^{\dag} b_{+} + b_{+}^{\dag} a_{-} )  
	+\mathrm{h.c.}
	\end{align}
	The full model including the dissipation and detuning terms is provided in the Supplement.

	We can now derive the equations of motion for this system by means of the Heisenberg-Langevin equation (see Supplementary Note 2). The key features of our data can be understood simply by a series of adiabatic elimination steps. First, we adiabatically eliminate the $c_+$ optical mode with linewidth $\kappa \gg \gamma_m$, which leads to sideband cooling of both $a_+$ and $b_+$. In particular, the mode $b_+$ having bare linewidth 
	$\mathit{\Gamma}$, is damped optically with a rate $\mathit{\Gamma} C_{\alpha}$ from this adiabatic elimination. The parameter $\mathcal{C}_\alpha$ is the quasi-mode optomechanical cooperativity defined as $4\alpha^2 g_0^2/\mathit{\Gamma}\kappa$.

	Viewing the quasi-mode as a bath, we see that this cw bath is cooled to a temperature $T_{b+} \approx T_{\rm bulk}/(1 + C_\alpha)$ via sideband cooling. This chiral refrigeration in turn modifies the damping and temperature of the $a_-$ mode. Specifically, adiabatic elimination of the quasi-mode $b_+$ leads to an effective damping of the $a_-$ mode. At zero optical power, we have bare linewidth $\gamma_m = \gamma + 4 |V_0|^2/\mathit{\Gamma}$, where $\gamma$ is the intrinsic linewidth of $a_-$.
	As we increase the optical power, we see that the disorder-induced damping term reduces due to the increased damping of the $b_+$ quasi-mode -- this is an optically induced impedance mismatch. Consequently, if there were no probe light, we would see that the $a_-$ damping rate reduces to 
	\begin{equation}
	\gamma_{a_-} = \gamma + \frac{4 |V_0|^2}{\mathit{\Gamma} (1 + C_\alpha)} ~. 
	\label{eq:linewidth}
	\end{equation}
	We can also see that the temperature of the $a_-$ mode should go to a weighted sum of these two terms (details in Supplementary Note 2):
	\begin{equation}
	T_{a-} = \frac{\gamma T_{\rm bulk} + (\gamma_{a_-} - \gamma) T_{b+}}{\gamma_{a_-}}
	\label{eq:temperature}
	\end{equation}
	As $T_{b+} < T_{\rm bulk}$, we see that for moderate $C_\alpha$, the temperature $T_{a-}$ goes down, even as $\gamma_{a_-} < \gamma_m$!
	Conventional optomechanical cooling involves an increase of mechanical damping while the heat load remains constant, resulting in a lowering of the mode temperature. Here, we see that while the damping reduces, the heat load on the system also reduces. This leads to a lower effective temperature of the mechanical system even as the linewidth narrows. 
	Important corrections due to the finite probe power lead to additional broadening and cooling of the $a_-$ mode, while the $a_+$ mode's dynamics are dominated by the sideband cooling from the $\alpha h_0$ coupling. But the key features are described by the above picture of chiral refrigeration.

	We excluded a simpler model, of two degenerate mechanical modes $a_\pm$ coupled by disorder of strength $V_1$ and no additional quasi-modes, as it fails to produce two key features of the data below. First, at low pump power, we would see significant mode splitting (below we set an experimental bound for the direct coupling rate between $a_\pm$ modes at $V_1 < 1$ kHz), representing a breaking of parity conservation due to disorder-induced scattering. Second, at high pump powers, explored below, the smallest linewidth that the backward mode could achieve would be equivalent to its initial linewidth and its temperature would be equal to the bath temperature. We present a more detailed analysis of this model in Supplementary Note 2. Optical coupling to multiple (bulk) mechanical modes is the next best alternative, and as we show above, describes these phenomena.

	\subsection{Disorder suppression and optomechanical cooling without damping}

	We now return to the experimental results to demonstrate the key predictions of this model: (i) damping associated with disorder-induced scattering can be optically inhibited, (ii) the damping rate of ccw phonons can be brought below the bare linewidth $\gamma_m$, and (iii) the process leads to a reduction in heat load. Here, we employ an erbium doped fiber amplifier to control the cw pump power ($P_{cw}$) while keeping the ccw probe power constant at 12.5 $\mu \text{W}$. The anti-Stokes Brillouin scattered light in the resonator is kept close to zero detuning from its optical mode to maximize cooling efficiency, i.e. $|\mathit{\Delta}_2 /\kappa|$ is always less than 10 \%. Since the ccw probe adds some fixed optical damping to the ccw phonons, the initial measurement of \gam is at $17.8(\pm 1)$ kHz, which is greater than the bare linewidth $\gamma_m$.

	\begin{figure}[hp!]
		\begin{adjustwidth}{-1in}{-1in}
			\centering
			\includegraphics[width=1.05\textwidth]{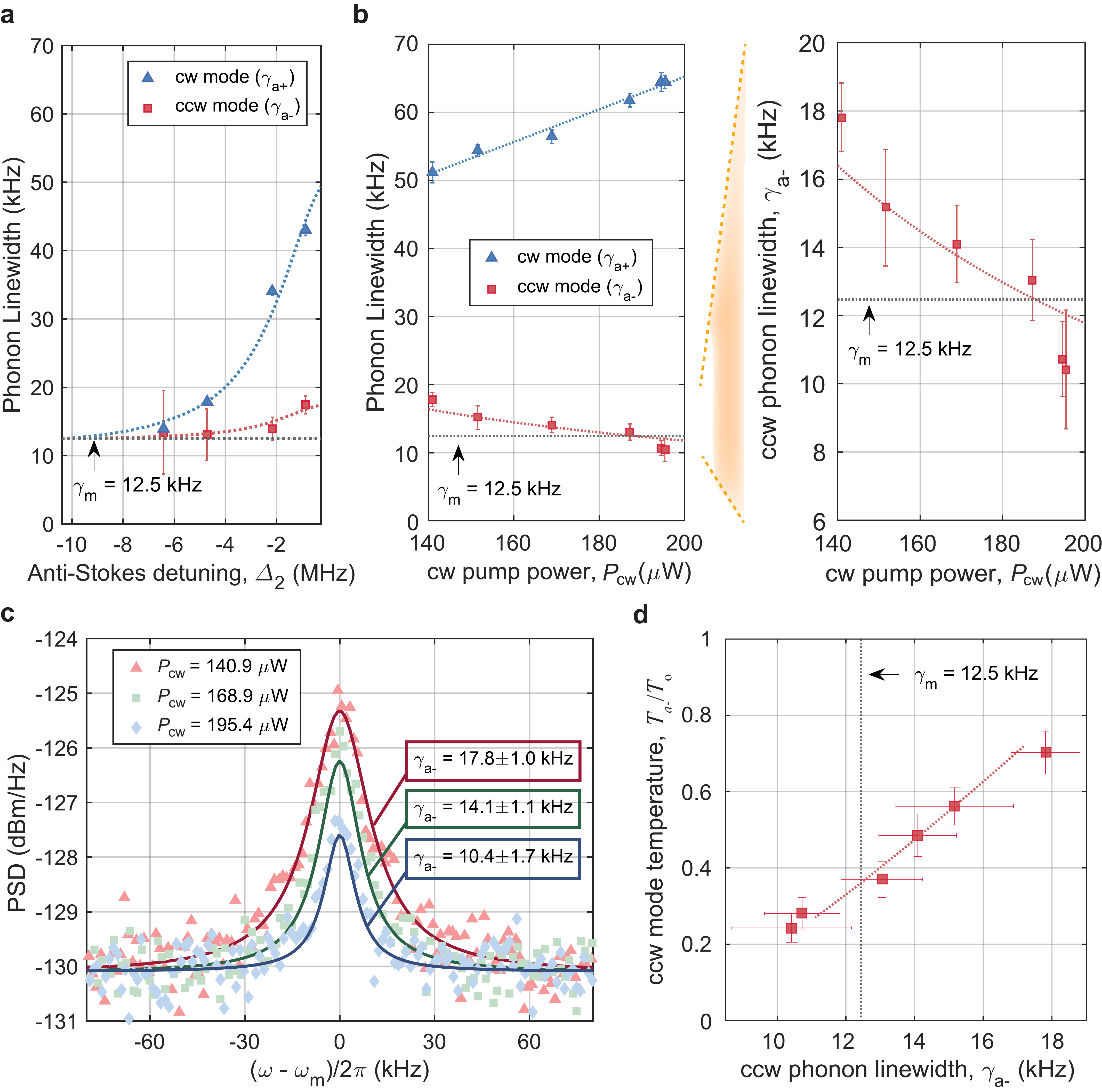}
			\caption{
				\textbf{Chiral refrigeration in a silica whispering gallery resonator.} 
				\textbf{(a)} Experimental measurement of bare phonon linewidth $\gamma_m$ and observation of chiral asymmetry in dissipation rates for cw and ccw phonons during the initial sideband cooling experiment. Blue and red dashed lines in subfigures a and b are fits to theory in Supplementary Eqns.~5.
				\textbf{(b)} Observation of increasing ccw phonon coherence during sideband cooling of the cw propagating phonons (fixed ccw probe). The ccw phonon linewidth \gam drops below the bare linewidth $\gamma_m$, indicating reduction in the disorder induced dissipation.
				\textbf{(c)} Measured photocurrent PSD (proportional to ccw phonon PSD) during experiment of subfigure b. The ccw phonons experience gain-free spectral narrowing and cooling (reduction in spectrum area) when the cw laser $P_\textrm{cw}$ is increased. Solid lines are Lorentzian fits to the data.
				\textbf{(d)} Measured ccw phonon mode temperature $T_{a_-}^{\text{eff}}$ vs linewidth \gam showing that, in contrast to sideband cooling, both the linewidth and temperature decrease during this experiment. Temperature is calculated using the integrated phonon power spectrum (examples in subfigure c), with vertical error bars generated from amplitude uncertainty. Red dashed line is a fit to the theory of Eqns.~\ref{eq:linewidth} and \ref{eq:temperature}. The phonon mode starts pre-cooled due to the sideband cooling from the probe.
				All error bars correspond to 95\,\% confidence intervals for their respective parameters.
			}
			\label{fig:main}
		\end{adjustwidth}
	\end{figure}

	As we increase cw pump power from 140.9 $\mu \text{W}$ to 195.4 $\mu \text{W}$, the added optical damping broadens the cw phonon linewidth \gap (Fig.~\ref{fig:main}b). The striking feature of this experiment is that the ccw phonon linewidth \gam simultaneously reduces, i.e. the ccw phonons become more coherent! We verify that the increased coherence of the ccw phonons is not associated with any gain (Fig.~\ref{fig:main}c) by observing their spectrum through the measured photocurrent (Supplementary Note 4). In fact, quite the opposite occurs, and the total integrated area under the phonon spectrum also reduces, indicating a reduction in temperature of the \am phonons. 
	Since the ccw optical probe was not modified, the optically induced damping from the probe laser remains fixed. The reduction of the \gam linewidth thus indicates that a hidden contribution to dissipation is being eliminated when the cw pump power is increased. At the highest power, the smallest dressed linewidth $\gam = 10.5 (\pm 1.7)$ kHz is below the bare linewidth $\gamma_m = 12.5 (\pm 1)$ kHz measured at the start of the experiment, even including the extra sideband damping from probe $\beta$. 
	Fitting of the power vs linewidth measurements to our model (Supplementary Eqns.~5) reveals the ratio of coupling rates $V_0/g_0 = 1.15 (\pm 0.05) \times 10^3$. 
	Our model indicates that the observed reduction in the phonon linewidth \gam occurs due to reduction of the disorder-induced scattering. Specifically, the ccw propagating phonons achieve appreciable robustness against disorder due to chiral optomechanical damping of the cw phonon quasi-mode.

	In Fig.~\ref{fig:main}d we present the temperature $T_{a-}$ of the ccw phonon mode measured through phonon power spectral area, as a function of its measured linewidth. 
	Fitting this temperature data to our model in Supplementary Note 2 permits extraction of $V_0^2 / \mathit{\Gamma} = 2.8 (\pm 0.32)$ kHz. 
	The parameters $V_0$, $g_0$, and $\mathit{\Gamma}$ cannot presently be further separated since the phonon quasi-modes $b_\pm$ are not directly observable. However, we note that the minimum self-consistent quasi-mode linewidth is approximately the optical linewidth ($\mathit{\Gamma} \approx \kappa$) and we can obtain the values $V_0 = 121 (\pm 13)$ kHz and $g_0 = 105 (\pm 12)$ Hz at this minimum. These estimates are commensurate with our earlier assumption that the disorder induced scattering between the high-$Q$ and quasi-modes dominates over direct scattering between the high-$Q$ modes (i.e. $V_0 \gg V_1$). We additionally learn that the lower limit of $g_0$ is roughly 7.5 times $h_0$, implying that the optomechanical coupling to the quasi-mode is significant, and which agrees with the number of phonon modes that are likely to compose the quasi-mode. The anomalous cooling that we observe is thus well explained by significant coupling to, and chiral refrigeration of the cw quasi-mode bath.

	\section{Discussion}
	Sideband cooling has been to date the only mechanism available for suppressing the thermal motion of mechanical resonators using light -- but is necessarily accompanied by linewidth broadening. In this work, we have demonstrated the existence of a fundamentally different mechanism for cooling mechanical oscillators, that occurs through sideband cooling of the bath modes. No previous experiment in optomechanics has provided either direct or indirect evidence of such bath cooling. More importantly, this mechanism has the potential to revolutionize the noise calculus that we employ, since the cooling is instead accompanied by linewidth narrowing! Additionally, we have demonstrated for the first time that not only can phonon chirality be induced optically, but also that it mitigates the influence of disorder on propagating phonons, a technique that potentially revolutionizes phonon-assisted measurements. To date such scattering immunity for phonons has only been demonstrated in topological insulators. Our results thus dramatically push forward the known physics for both laser cooling and for monolithic chiral systems.
	
	Our approach for inducing chiral behavior is, at present, confined to the narrowband response of a high-$Q$ resonator system. However, such devices are already in use for metrological applications \cite{Teufel:2009gn,PhysRevA.82.061804,Krause:2012cf,Gavartin:2012el} including atomic force microscopes \cite{Aksyuk12} and quantum-regime transducers \cite{lehnert15,cleland13}. In all these cases, increasing the quality factor while reducing the heat load of the mechanical element would lead to a direct improvement in performance. Furthermore, the modification of phonon transport by light may have substantial impact even beyond contemporary devices, as the ability to dynamically reconfigure the phononic behavior may change the realm of possibility as currently conceived. Still, robust demonstration of chiral asymmetry and non-reciprocal behavior remains close, and our work provides a foundation upon which to build such demonstrations.

	\section*{Data availability}
	Data can be made available by request to the authors on an individual basis. 
	
	\section*{Acknowledgements}
	Funding for this research was provided through the National Science Foundation (NSF), Air Force Office for Scientific Research (AFOSR), the Office of Naval Research (ONR), and DARPA MTO.
	
	\section*{Author contributions}
	SK, JMT and GB conceived and designed the experiments. 
	SK developed the experimental setup and carried out the experiments. 
	XX and JMT developed the theoretical analysis. 
	All authors jointly analyzed the data and co-wrote the paper. 
	JMT and GB supervised all aspects of this project.

	\def\url#1{}
	\bibliographystyle{ieeetr}

\begin{thebibliography}{10}
		
		\bibitem{Kippenberg555}
		T.~J. Kippenberg, R.~Holzwarth, and S.~A. Diddams, ``Microresonator-based
		optical frequency combs,'' {\em Science}, vol.~332, no.~6029, pp.~555--559,
		2011.
		
		\bibitem{Li309}
		J.~Li, X.~Yi, H.~Lee, S.~A. Diddams, and K.~J. Vahala, ``Electro-optical
		frequency division and stable microwave synthesis,'' {\em Science}, vol.~345,
		no.~6194, pp.~309--313, 2014.
		
		\bibitem{Loh:15}
		W.~Loh, A.~A.~S. Green, F.~N. Baynes, D.~C. Cole, F.~J. Quinlan, H.~Lee, K.~J.
		Vahala, S.~B. Papp, and S.~A. Diddams, ``Dual-microcavity narrow-linewidth
		brillouin laser,'' {\em Optica}, vol.~2, pp.~225--232, Mar 2015.
		
		\bibitem{Gigan:2006p1091}
		S.~Gigan, H.~Bohm, M.~Paternostro, F.~Blaser, G.~Langer, J.~Hertzberg,
		K.~Schwab, D.~Bauerle, M.~Aspelmeyer, and A.~Zeilinger, ``Self-cooling of a
		micromirror by radiation pressure,'' {\em Nature}, vol.~444, pp.~67--70, Nov
		2006.
		
		\bibitem{Arcizet:2006hv}
		O.~Arcizet, P.-F. Cohadon, T.~Briant, M.~Pinard, and A.~Heidmann,
		``Radiation-pressure cooling and optomechanical instability of a
		micromirror,'' {\em Nature}, vol.~444, pp.~71--74, Nov 2006.
		
		\bibitem{Chan_GroundState_2011}
		J.~Chan, T.~M. Alegre, A.~H. Safavi-Naeini, J.~T. Hill, A.~Krause,
		S.~Gr{\"o}blacher, M.~Aspelmeyer, and O.~Painter, ``{Laser cooling of a
			nanomechanical oscillator into its quantum ground state},'' {\em Nature},
		vol.~478, no.~7367, pp.~89--92, 2011.
		
		\bibitem{Verhagen:2012ei}
		E.~Verhagen, S.~Deleglise, S.~Weis, A.~Schliesser, and T.~J. Kippenberg,
		``{Quantum-coherent coupling of a mechanical oscillator to an optical cavity
			mode},'' {\em Nature}, vol.~482, pp.~63--67, Feb. 2012.
		
		\bibitem{Teufel:2009gn}
		J.~D. Teufel, T.~Donner, M.~A. Castellanos-Beltran, J.~W. Harlow, and K.~W.
		Lehnert, ``{Nanomechanical motion measured with an imprecision below that at
			the standard quantum limit},'' {\em Nat. Nanotech.}, vol.~4, pp.~820--823,
		Nov. 2009.
		
		\bibitem{PhysRevA.82.061804}
		G.~Anetsberger, E.~Gavartin, O.~Arcizet, Q.~P. Unterreithmeier, E.~M. Weig,
		M.~L. Gorodetsky, J.~P. Kotthaus, and T.~J. Kippenberg, ``Measuring
		nanomechanical motion with an imprecision below the standard quantum limit,''
		{\em Phys. Rev. A}, vol.~82, p.~061804, Dec 2010.
		
		\bibitem{Krause:2012cf}
		A.~G. Krause, M.~Winger, T.~D. Blasius, Q.~Lin, and O.~Painter, ``{A
			high-resolution microchip optomechanical accelerometer},'' {\em Nat.
			Photonics}, vol.~6, pp.~768--772, Oct. 2012.
		
		\bibitem{Gavartin:2012el}
		E.~Gavartin, P.~Verlot, and T.~J. Kippenberg, ``{A hybrid on-chip
			optomechanical transducer for ultrasensitive force measurements},'' {\em Nat.
			Nanotech.}, vol.~7, pp.~509--514, June 2012.
		
		\bibitem{Halperin:1982tb}
		B.~I. Halperin, ``{Quantized Hall Conductance, Current-Carrying Edge States,
			and the Existence of Extended States in a Two-Dimensional Disordered
			Potential},'' {\em Phys. Rev. B}, vol.~25, no.~4, pp.~2185--2190, 1982.
		
		\bibitem{Wang09}
		Z.~Wang, Y.~Chong, J.~D. Joannopoulos, and M.~Soljacic, ``{Observation of
			unidirectional backscattering-immune topological electromagnetic states},''
		{\em Nature}, vol.~461, pp.~772--775, Aug. 2009.
		
		\bibitem{Hafezi:2013jg}
		M.~Hafezi, S.~Mittal, J.~Fan, A.~Migdall, and J.~M. Taylor, ``{Imaging
			topological edge states in silicon photonics},'' {\em Nat. Photonics},
		vol.~7, pp.~1001--1005, Oct. 2013.
		
		\bibitem{Susstrunk15}
		R.~S{\"u}sstrunk and S.~D. Huber, ``{Observation of phononic helical edge
			states in a mechanical topological insulator},'' {\em Science}, vol.~349,
		no.~6243, pp.~47--50, 2015.
		
		\bibitem{Peano:2016je}
		V.~Peano, M.~Houde, C.~Brendel, F.~Marquardt, and A.~A. Clerk, ``{Topological
			phase transitions and chiral inelastic transport induced by the squeezing of
			light},'' {\em Nat. Commun.}, vol.~7, Mar. 2016.
		
		\bibitem{Fleury:2014fn}
		R.~Fleury, D.~L. Sounas, C.~F. Sieck, M.~R. Haberman, and A.~Alu, ``{Sound
			Isolation and Giant Linear Nonreciprocity in a Compact Acoustic
			Circulator},'' {\em Science}, vol.~343, pp.~516--519, Jan. 2014.
		
		\bibitem{Estep:2014jp}
		N.~A. Estep, D.~L. Sounas, J.~Soric, and A.~Alu, ``{Magnetic-free
			non-reciprocity and isolation based on parametrically modulated
			coupled-resonator loops},'' {\em Nat. Phys.}, vol.~10, pp.~923--927, Nov.
		2014.
		
		\bibitem{Kim2015}
		J.~Kim, M.~C. Kuzyk, K.~Han, H.~Wang, and G.~Bahl, ``{Non-reciprocal Brillouin
			scattering induced transparency},'' {\em Nat. Phys.}, vol.~11, pp.~275--280,
		Mar. 2015.
		
		\bibitem{Dong2015}
		C.-H. Dong, Z.~Shen, C.-L. Zou, Y.-L. Zhang, W.~Fu, and G.-C. Guo,
		``Brillouin-scattering-induced transparency and non-reciprocal light
		storage,'' {\em Nat. Commun.}, vol.~6, Feb. 2015.
		
		\bibitem{Shen:gt}
		Z.~Shen, Y.-L. Zhang, Y.~Chen, C.-L. Zou, Y.-F. Xiao, X.-B. Zou, F.-W. Sun,
		G.-C. Guo, and C.-H. Dong, ``{Experimental realization of optomechanically
			induced non-reciprocity},'' {\em Nat Photon}, vol.~10, no.~10, pp.~657--661,
		2016.
		
		\bibitem{XuTaylor_2016}
		X.~Xu, T.~Purdy, and J.~M. Taylor, ``Cooling a harmonic oscillator by
		optomechanical modification of its bath,'' {\em arXiv.org}, Aug 2016.
		\newblock arXiv:1608.05717.
		
		\bibitem{Gorodetsky:2000vr}
		M.~L. Gorodetsky, A.~D. Pryamikov, and V.~S. Ilchenko, ``{Rayleigh scattering
			in high-Q microspheres},'' {\em Journal of the Optical Society of America B},
		vol.~17, pp.~1051--1057, June 2000.
		
		\bibitem{Knopoff:1964ve}
		L.~Knopoff and J.~A. Hudson, ``{Scattering of elastic waves by small
			inhomogeneities},'' {\em The Journal of the Acoustical Society of America},
		vol.~36, pp.~338--343, Feb. 1964.
		
		\bibitem{Pao:1976wc}
		Y.~H. Pao and V.~Varatharajulu, ``{Huygens Principle, Radiation Conditions, and
			Integral Formulas for Scattering of Elastic-Waves},'' {\em Journal of the
			Acoustical Society of America}, vol.~59, no.~6, pp.~1361--1371, 1976.
		
		\bibitem{Bahl:2012jm}
		G.~Bahl, M.~Tomes, F.~Marquardt, and T.~Carmon, ``{Observation of spontaneous
			Brillouin cooling},'' {\em Nat. Phys.}, vol.~8, pp.~203--207, Mar. 2012.
		
		\bibitem{marquardt.2008}
		F.~Marquardt, A.~A. Clerk, and S.~M. Girvin, ``{Quantum theory of
			optomechanical cooling},'' {\em J. Mod. Opt.}, vol.~55, no.~19-20,
		pp.~3329--3338, 2008.
		
		\bibitem{park.2009}
		Y.-S. Park and H.~Wang, ``{Resolved-sideband and cryogenic cooling of an
			optomechanical resonator},'' {\em Nat. Phys.}, vol.~5, no.~7, pp.~489--493,
		2009.
		
		\bibitem{Bahl2011a}
		G.~Bahl, J.~Zehnpfennig, M.~Tomes, and T.~Carmon, ``{Stimulated optomechanical
			excitation of surface acoustic waves in a microdevice},'' {\em Nat. Commun.},
		vol.~2, pp.~403--406, 2011.
		
		\bibitem{QuasiModeTheory_2016}
		X.~Xu, S.~Kim, G.~Bahl, and J.~M. Taylor, ``A quasi-mode theory of chiral
		phonons,'' {\em arXiv.org}, Dec 2016.
		\newblock arXiv:1612.09240.
		
		\bibitem{Aksyuk12}
		Y.~Liu, H.~Miao, V.~Aksyuk, and K.~Srinivasan, ``Wide cantilever stiffness
		range cavity optomechanical sensors for atomic force microscopy,'' {\em Opt.
			Express}, vol.~20, pp.~18268--18280, Jul 2012.
		
		\bibitem{lehnert15}
		R.~W. Andrews, A.~P. Reed, K.~Cicak, J.~D. Teufel, and K.~W. Lehnert,
		``Quantum-enabled temporal and spectral mode conversion of microwave
		signals,'' {\em Nat. Commun.}, vol.~6, 11 2015.
		
		\bibitem{cleland13}
		J.~Bochmann, A.~Vainsencher, D.~D. Awschalom, and A.~N. Cleland,
		``Nanomechanical coupling between microwave and optical photons,'' {\em Nat.
			Phys.}, vol.~9, pp.~712--716, 11 2013.
		
	\end{thebibliography}

\begin{thebibliography}{}
		\providecommand{\url}[1]{#1}
		\csname url@samestyle\endcsname
		\providecommand{\newblock}{\relax}
		\providecommand{\bibinfo}[2]{#2}
		\providecommand{\BIBentrySTDinterwordspacing}{\spaceskip=0pt\relax}
		\providecommand{\BIBentryALTinterwordstretchfactor}{4}
		\providecommand{\BIBentryALTinterwordspacing}{\spaceskip=\fontdimen2\font plus
			\BIBentryALTinterwordstretchfactor\fontdimen3\font minus
			\fontdimen4\font\relax}
		\providecommand{\BIBforeignlanguage}[2]{{%
				\expandafter\ifx\csname l@#1\endcsname\relax
				\typeout{** WARNING: IEEEtran.bst: No hyphenation pattern has been}%
				\typeout{** loaded for the language `#1'. Using the pattern for}%
				\typeout{** the default language instead.}%
				\else
				\language=\csname l@#1\endcsname
				\fi
				#2}}
		\providecommand{\BIBdecl}{\relax}
		\BIBdecl
		
		\bibitem{xunnongCooling16}
		X.~Xu, T.~Purdy, and J.~M. Taylor, ``Cooling a harmonic oscillator by
		optomechanical modification of its bath,'' \emph{arXiv.org}, Aug 2016,
		arXiv:1608.05717.
		
		\bibitem{Kim2015}
		J.~Kim, M.~C. Kuzyk, K.~Han, H.~Wang, and G.~Bahl, ``{Non-reciprocal Brillouin
			scattering induced transparency},'' \emph{Nature Physics}, vol.~11, no.~3,
		pp. 275--280, Mar. 2015.
		
		\bibitem{gardiner2004quantum}
		C.~Gardiner and P.~Zoller, \emph{Quantum noise: a handbook of Markovian and
			non-Markovian quantum stochastic methods with applications to quantum
			optics}.\hskip 1em plus 0.5em minus 0.4em\relax Springer Science \& Business
		Media, 2004.
		
		\bibitem{Safavi-NaeiniPainter2011}
		\BIBentryALTinterwordspacing
		A.~H. Safavi-Naeini and O.~Painter, ``Proposal for an optomechanical traveling
		wave phonon-photon translator,'' \emph{New Journal of Physics}, vol.~13,
		no.~1, p. 013017, 2011.
		\BIBentrySTDinterwordspacing
		
	\end{thebibliography}

\clearpage

\beginsupplement

\begin{center}
	\textbf{\large Supplementary Table 1. Table of Symbols}\\
\end{center}
{\renewcommand{\arraystretch}{1}
	\begin{table}[ht]
		\centering
		\begin{adjustbox}{width=1\textwidth}
			\begin{tabular}{ | c| c  | } 
				\hline
				\hline
				symbol & meaning\\ 
				\hline
				$c_{p, \sigma}$, $c_\sigma$& Two frequency-adjacent optical modes; the pump mode ($c_{p, \sigma}$) and the anti-Stokes mode ($c_\sigma$)\\  
				$c_{+(-)}$& Annihilation operator of the optical mode in the cw(ccw) direction\\ 
				$a_{+(-)}$& Annihilation operator of the high-$Q$ phonon mode in the cw(ccw) direction \\ 
				$b_k$& Annihilation operator of the phonon modes in the system\\ 
				$\lambda_k$& Optomechanical single-photon coupling strengths to the two optical modes; the $c_\sigma$ and $c_{p, \sigma}$ modes\\
				$\Lambda_k$& Pump-enhanced optomechanical coupling constant, $\Lambda_k \triangleq |\alpha \lambda_k|$. \\
				$\mu_k$& Coupling strength of disorder-induced scattering between the $a_{\pm}$ modes to the $b_k$ modes\\ 
				$b_{+(-)}$& Annihilation operator of the phonon quasi-mode in the cw(ccw) direction \\ 
				$c_{+(-)}^{\mr{in}}$& Annihilation operator of the optical noise in the $c_{+(-)}$ modes \\
				$a_-^{\mr{in}}$, $b_+^{\mr{in}}$ & Annihilation operator of the thermal noise in the $a_-$ and $b_+$ modes, respectively \\
				$a_-^{\mr{eff}}$ & Annihilation operator of the effective thermal noise in the $a_-$ mode\\
				$h_0$, $g_0$& Optomechanical single-photon coupling strengths, $h_0^2 + g_0^2 =1$\\
				$V_0$& Coupling strength of disorder-induced scattering between the $a_{\pm}$ modes to the $b_\mp$ mode\\ 
				$\omega_{1}$& Cavity resonance frequency of the pump optical mode \\ 
				$\omega_{2}$& Cavity resonance frequency of the anti-Stokes optical mode\\ 
				$\omega_L$& Pump laser frequency \\ 
				$\omega_m$& Mechanical resonance frequency \\ 
				$\delta$& Detuning of the pump mode from the pump laser, $\delta  = \omega_{1} - \omega_L $ \\ 
				$\mathit{\Delta}$& Detuning of the anti-Stokes mode from the pump laser, $\mathit{\Delta}  = \omega_{2} - \omega_L$ \\ 
				$\mathit{\Delta_2}$& Detuning of the anti-Stokes mode from the scattered light, $\mathit{\Delta_2}  = \omega_{2} - (\omega_L + \omega_m)$ \\ 
				$\kappa_{0}$& Intrinsic loss rate of the anti-Stokes optical mode \\ 
				$\kappa_{ex}$& Loss rate associated with the external coupling  \\ 
				$\kappa$& Measurable optical linewidth of the anti-Stokes mode, $\kappa = \kappa_0 + \kappa_{ex}$ \\
				$\kappa_{\mr{p}}$& Measurable optical linewidth of the pump mode \\
				$\gamma$& Mechanical damping rate of the high-$Q$ phonon modes\\ 
				$\mathit{\Gamma}$& Mechanical damping rate of the phonon quasi-modes\\ 
				$n_+ = |\alpha|^2$& Intracavity photon number in the cw optical mode $c_{+}$\\ 
				$n_- = |\beta|^2$& Intracavity photon number in the ccw optical mode $c_{-}$\\ 
				$\mathcal{C}_\alpha$& Cooperativity of the $c_+$ mode,  $\mathcal{C}_\alpha = 4\alpha^2 g_0^2/\mathit{\Gamma}\kappa$\\
				$\mathcal{C}_\beta$& Cooperativity of the $c_-$ mode, $\mathcal{C}_\beta =  4\beta^2 g_0^2/\mathit{\Gamma}\kappa$ \\
				$\chi_{a_\pm}^{-1} (\omega)$& Mechanical susceptibilities of the $a_{\pm}$ modes \\
				$T_{a_\pm}$& Bath temperatures of the $a_{\pm}$ modes \\	
				$T_{a_\pm}^{\mr{eff}}$& Effective temperatures of the $a_{\pm}$ modes \\		
				$n_L$& Intracavity photon number in the pump optical mode driven by the pump laser \\	
				$\bar{n}$& Effective phonon occupation number of the $a_-$ phonon mode \\
				$S_{a_{-}}(\omega)$& Quantum noise spectrum of $a_-$ mode, $S_{a_{-}}(\omega) = \int_{-\infty}^{\infty}dt e^{i\omega t}\langle a_-(t)a_-(0)  \rangle$\\	
				$S_{II}(\omega)$& Quantum noise spectrum of normalized photocurrent $I(t)$, $S_{II}(\omega) = \int_{-\infty}^{\infty}dt e^{i\omega t}\langle I(t)I(0)  \rangle$\\	
				\hline
				\hline
			\end{tabular}
		\end{adjustbox}
	\end{table}

	\newpage

	%
	\begin{center}
		\textbf{\large Supplementary Note 1. Defining the modes and their coupling}
	\end{center}
	
	Our system is composed of cw and ccw optical modes, high-$Q$ phonon modes, as well as vibrational excitations inside the material, i.e. the phonon bath. 
	As described in the  main text, we focus on a scenario in which high-$Q$ modes  of clockwise and counterclockwise circulation with annihilation operators  $a_\sigma$ are coupled via disorder to a quasi-mode (broad mode representing many actual mechanical modes) circulating in the opposite direction with annihilation operators $b_{\bar \sigma}$. 
	{In the experiment, for each circulation we have a pair of optical modes $c_{p, \sigma}$ and $c_\sigma$. The optomechanical coupling allows for transfer of light from  $c_{p, \sigma}$ to  $c_{\sigma}$ with a corresponding annihilation of a phonon that is phase matched. This process overlaps with both the high-$Q$ modes and the quasi-modes.}

	With the above basic picture, we examine this model using the rotating wave approximation (RWA) as the experimental configuration is all narrowband. We can then use the input-operator language to describe the open system dynamics. 
	{After displacing the optical cavity fields by the pump amplitudes in the cavity, $c_\sigma \rightarrow \sqrt{n_\sigma} + c_\sigma$ with $\sqrt{n_{+}} = \alpha, \sqrt{n_-} = \beta$, the $c_{p,\sigma}$ fluctuations decouple from the rest of the system.}
	Working in the frame rotating with the pump laser frequency, we write the linearized Heisenberg-Langevin equations for the mechanical and optical modes in the Fourier domain with Fourier frequency $\omega$:
	\begin{subequations}
		\begin{align}
		-i\omega c_\sigma &= -i \mathit{\Delta} c_\sigma - {\frac{\kappa}{2}} c_\sigma + \sqrt{\kappa} c_{_\sigma}^{\mr{in}} - i \sqrt{n_{\sigma}} (g_0 b_\sigma 	+ h_0 a_\sigma) 	\label{eq:1a}\\
		-i\omega a_\sigma &= -i \omega_m a_\sigma - {\frac{\gamma}{2}} a_\sigma + \sqrt{\gamma} a^{\mr{in}}_\sigma - i V_0 b_{\bar \sigma} - i h_0 \sqrt{n_\sigma} c_\sigma 	\label{eq:1b}\\
		-i\omega b_\sigma &= -i \omega_b b_\sigma  - {\frac{\mathit{\Gamma}}{2}} b_\sigma  +\sqrt{\mathit{\Gamma}} b^{\mr{in}}_\sigma  - i \alpha h_0 c_\sigma - i V_0 a_{\bar \sigma}
		\label{eq:1c}
		\end{align}
		\label{eq:1}
	\end{subequations} 
	where $V_0$ is the coupling strength of disorder-induced scattering between the cw(ccw) high-$Q$ phonon modes 
	$a_\sigma$ to the ccw(cw) phonon quasi-modes 
	$b_{\bar\sigma}$.
	{In contrast to the usual quantum optics literature, we here define detuning to be the mode frequency minus the signal frequency. Thus a positive detuning is red detuned. This makes comparison to mechanical motion as transparent as possible.}
	Meanwhile, the coupling of the quasi-modes $b_\sigma$ to the $c_\sigma$
	modes is given by $g_0 \sqrt{n_\sigma}$, while $h_0 \sqrt{n_{\sigma}}$ captures the coupling between $a_\sigma$ and $c_\sigma$.
	For any given $\omega$, there is a self-consistent $\omega_b \approx \omega$ that describes the relevant portion of the bath modes. Thus we take $|\omega_b - \omega| \ll \mathit{\Gamma}$ in what follows.
	Specifically, we have the following main assumptions for this simple model: 
	\begin{enumerate}
		\item Phonon backscattering occurs between high-$Q$ phonon modes and the phonon quasi-modes, i.e. between $a_{+} \longleftrightarrow b_{-}$ and $a_{-}  \longleftrightarrow b_{+}$, with strength $V_0 $. 
		\item The cw(ccw) optical mode $c_{+(-)}$ couples to the high-$Q$ phonon mode $a_{+(-)}$ and the cw(ccw) phonon quasi-mode $b_{+(-)}$ with different strengths. The cw optical mode $c_+$ couples to the cw high-$Q$ mode $a_+$ via direct optomechanical interaction with strength $\alpha h_0$ and couples to the quasi-mode with strength $\alpha g_0$. Likewise, the ccw optical mode $c_-$ couples to the ccw high-$Q$ mode $a_-$ with strength $ \beta h_0$ and couples to the ccw quasi-mode $b_-$ with strength $\beta g_0 $. Here are $n_+$ and $n_-$ are the number of intracavity photon in the cw optical mode $c_+$ and the ccw mode $c_-$, respectively.
		\item The high-$Q$ phonon modes $a_{+(-)}$ and the phonon quasi-modes $b_{+(-)}$ have the intrinsic damping rates $\gamma$ and $\mathit{\Gamma}$, respectively ($\gamma \ll \mathit{\Gamma}$). The cw modes and ccw modes have symmetry with respect to the origin. We also assume that the damping rate $\mathit{\Gamma}$ is in the same order as the optical loss rate $\kappa$. 
	\end{enumerate}
	We exclude a simpler model, of two degenerate mechanical modes and no additional quasi-modes, as it fails to produce two key features of the data. First, at low pump power, we would experimentally observe some mode splitting, representing a breaking of circular symmetry from disorder-induced scattering. Second, at high pump power, the lowest linewidth the backward mode could achieve would be equivalent to its initial linewidth, and its temperature would be equal to the bath temperature. Optical coupling to multiple mechanical modes is the next best alternative, and as we show here, describes these phenomena.

	Based on these assumptions, we can obtain the simplified continuum model as shown in Supplementary Fig.~\ref{fig:s1}.\textbf{a}. 
	In principle, the dynamics of the system can be solved numerically. However the loop structure in this coupled six-mode system will complicate the result, rendering interpretation difficult. To better capture the main physics, we can make the following approximation:
	we assume the $g_0$ parameter is larger than $h_0$ so that the optical field couples more strongly to the bulk modes $b_{\pm}$. 
	We note that this assumption is not actually that important for our main result, as the crucial point is that the dominant mechanical damping mechanism is coupling of the high-$Q$ modes to the quasi-modes -- this is independent of $g_0$ and $h_0$.
	We can then break the loop into two pieces (see Supplementary Fig.~\ref{fig:s1}.\textbf{a}-\textbf{c}).

	\begin{figure}[!hp]
		\begin{center}
			\includegraphics[width=.75\textwidth]{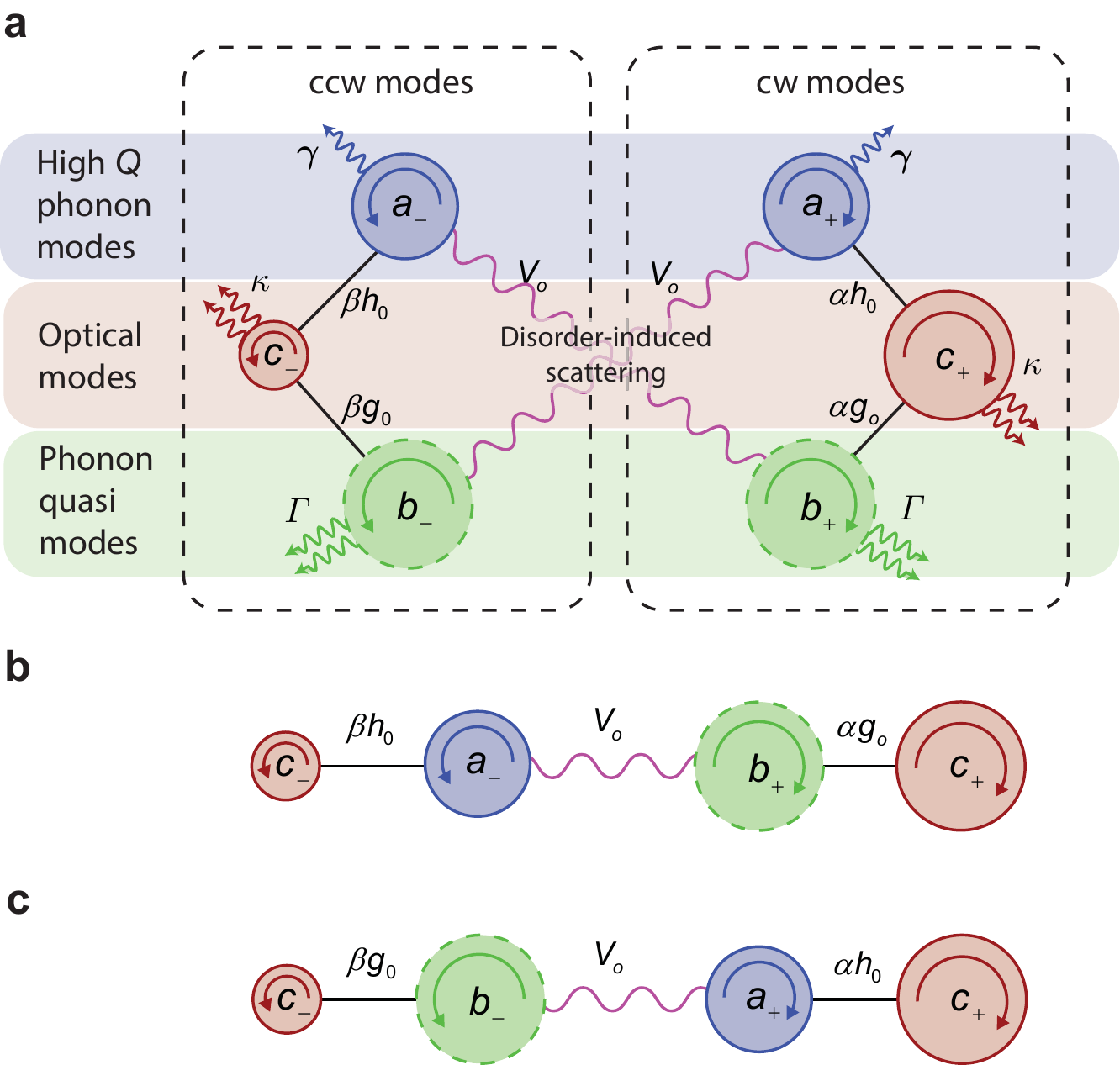}
			\caption{(a) Full description of the optomechanical coupling of the $c_{\pm}$ modes to $a_{\pm}$ modes and $b_\pm$ modes, and the disorder-induced scattering between the $a_{\pm}$ modes and $b_\mp$ modes. Separated optomechanical coupling descriptions by coupling directions (b)-(c). (b) The ccw-side coupling that the optical mode $c_-$ is mainly coupled to the $a_-$ mode with the disorder-induced scattering. (c) Likewise, the cw-side coupling that the cw mode $c_-$ is coupled to the $a_+$ mode. }
			\label{fig:s1}
		\end{center}
	\end{figure}
	
	\FloatBarrier

	\vspace{24pt}
	\begin{center}
		\textbf{\large Supplementary Note 2. Analysis of the simplified continuum model}
	\end{center}
	
	\vspace{24pt}
	
	\noindent
	\textbf{SN2.1 Susceptibilities of the high-$Q$ modes,} {$a_{\pm}$}
	\vspace{12pt}
	
	We first focus on the ccw-side coupling direction (Supplementary Figure~\ref{fig:1}.\textbf{b}) to calculate the linewidth of the ccw phonon $a_-$. 
	Solving in the Fourier domain, we get 
	\bea
	&& \left[ -i(\omega -\omega_m) + \gamma/2 + \frac{\beta^2 h_0^2}{-i(\omega - \mathit{\Delta}) + \kappa/2} +  \frac{V_0^2}{\mathit{\Gamma}/2} \left(1 - \frac{\alpha^2 g_0^2}{\mathit{\Gamma}\tilde{\kappa} /2}  \right)   \right] a_-  \nonumber \\
	&=&  \sqrt{\gamma}a_-^{\mr{in}}  -  \frac{i \beta h_0 \sqrt{\kappa} }{-i(\omega - \mathit{\Delta}) + \kappa/2} c_-^{\text{in}}  - \frac{V_0\alpha g_0 \sqrt{\kappa}}{\mathit{\Gamma}\tilde{\kappa} /2} c_+^{\text{in}} - \frac{iV_0}{\sqrt{\mathit{\Gamma}}/2} \left( 1 - \frac{\alpha^2 g_0^2}{\mathit{\Gamma}\tilde{\kappa}/2} \right) b_+^{\text{in}} \label{eq:8}
	\eea
	where $\tilde{\kappa} \triangleq -i(\omega - \mathit{\Delta}) + \kappa/2 + 2\alpha^2 g_0^2/\mathit{\Gamma}$. 
	Similarly, for the cw phonon mode $a_+$, we can find its equation of motion by interchanging $\alpha$ with $\beta$, $a_+$ with $a_-$, and $c_+$ with $c_-$: 
	\bea
	&& \left[ -i(\omega -\omega_m) + \gamma/2 + \frac{\alpha^2 h_0^2}{-i(\omega - \mathit{\Delta}) + \kappa/2} +  \frac{V_0^2}{\mathit{\Gamma}/2} \left(1 - \frac{\beta^2 g_0^2}{\mathit{\Gamma}\tilde{\kappa}^{\prime} /2}  \right)   \right] a_+  \nonumber \\
	&=&  \sqrt{\gamma}a_+^{\mr{in}}  -  \frac{i \alpha h_0 \sqrt{\kappa} }{-i(\omega - \mathit{\Delta}) + \kappa/2} c_+^{\text{in}}  - \frac{V_0\beta g_0 \sqrt{\kappa}}{\mathit{\Gamma}\tilde{\kappa}^{\prime} /2} c_-^{\text{in}} - \frac{iV_0}{\sqrt{\mathit{\Gamma}}/2} \left( 1 - \frac{\beta^2 g_0^2}{\mathit{\Gamma}\tilde{\kappa}^{\prime}/2} \right) b_-^{\text{in}} 
	\eea
	with $\tilde{\kappa}^{\prime} \triangleq -i(\omega - \mathit{\Delta}) + \kappa/2 + 2\beta^2 g_0^2/\mathit{\Gamma}$. 
	The susceptibilities of the $a_{\pm}$ modes are given by the left hand side of the equations of motion: 
	\begin{subequations}
		\bea
		\chi_{a_+}^{-1} (\omega) &=& -i(\omega -\omega_m) + \gamma/2 + \frac{\alpha^2h_0^2}{-i(\omega - \mathit{\Delta}) + \kappa/2} +  \frac{V_0^2}{\mathit{\Gamma}/2} \left(1 - \frac{\beta^2 g_0^2}{\mathit{\Gamma}\tilde{\kappa}^{\prime} /2}  \right) , \\
		\chi_{a_-}^{-1} (\omega) &=& -i(\omega -\omega_m) + \gamma/2 + \frac{\beta^2h_0^2}{-i(\omega - \mathit{\Delta}) + \kappa/2} +  \frac{V_0^2}{\mathit{\Gamma}/2} \left(1 - \frac{\alpha^2 g_0^2}{\mathit{\Gamma}\tilde{\kappa} /2}  \right). 
		\eea
	\end{subequations}

	\vspace{24pt}
	\noindent
	\textbf{SN2.2 Linewidths of the  $a_{\pm}$ modes}
	
	\vspace{12pt}
	We can define the cooperativities as $\mathcal{C}_\alpha = 4\alpha^2 g_0^2/\mathit{\Gamma}\kappa$ and $\mathcal{C}_\beta =  4\beta^2 g_0^2/\mathit{\Gamma}\kappa$, which are both dimensionless parameters describing the strength of optomechanical coupling relative to cavity decay rate and mechanical damping rate. 
	Under the phase matching condition that the pump laser and its scattered light are near the two frequency-adjacent optical modes, we can expect $\mathit{\Delta} \approx \omega_m$ (see Supplementary Fig.~\ref{fig:s3}). To evaluate the linewidth of the high-$Q$ phonon modes, we set $\omega \approx \omega_m$. Then we have:
	\begin{subequations}
		\bea
		\gamma_{a_+} &=& \gamma + \frac{4\alpha^2h_0^2}{\kappa} +  \frac{4V_0^2}{\mathit{\Gamma}}  \frac{\kappa}{ \kappa + 4\beta^2  g_0^2/\mathit{\Gamma} }  = \gamma + \frac{4\alpha^2h_0^2}{\kappa} +  \frac{4V_0^2}{\mathit{\Gamma}}  \frac{1}{ 1 + \mathcal{C}_\beta }    ,  \\
		\gamma_{a_-}  &=& \gamma + \frac{4\beta^2h_0^2}{\kappa} +  \frac{4V_0^2}{\mathit{\Gamma}}  \frac{\kappa}{ \kappa + 4\alpha^2 g_0^2/\mathit{\Gamma} } =  \gamma + \frac{4\beta^2h_0^2}{\kappa} +  \frac{4V_0^2}{\mathit{\Gamma}}   \frac{1}{ 1 + \mathcal{C}_\alpha }   
		\eea
	\end{subequations}
	where $\gamma_{a_+}$ and $\gamma_{a_-}$ are  the linewidths of the high-$Q$ phonon modes $a_{\pm}$. Note that the linewidths ${\gamma_{a_\pm}}$ are larger than their minimum measurable linewidth $\gamma_m {= \gamma + \frac{4V_0^2}{\mathit{\Gamma}}}$ {(obtained when optical power is zero, i.e. $\alpha = 0$, $\beta = 0$)} due to the disorder induced backscattering to the counter-propagating quasimode (see Eq.~2)

	\vspace{36pt}
	\noindent
	\textbf{SN2.3 Effective temperature of the phonon modes}
	\vspace{12pt}
	
	Another important feature that comes from the continuum model is the reduction in the effective temperature of the $a_-$ mode, because of coherent damping of the $\sigma = +$ mechanical modes. 
	When the right-hand side of the equation (\ref{eq:8}) is considered with an assumption that the optical noise $c_\sigma^{\mr{in}}$ is negligible compared to the thermal noise source, we have the effective noise on $a_-$ as:
	\bea
	\sqrt{\gamma} a_-^{\text{in}}  - \frac{iV_0}{\sqrt{\mathit{\Gamma}}/2} \left( 1 - \frac{\alpha^2 g_0^2}{\mathit{\Gamma}\tilde{\kappa}/2} \right) b_+^{\text{in}} \label{eq:18}
	\eea
	The effective temperature of mode $a_-$ is then given by:
	\bea
	T_{a_-}^ {\text{eff}} &=& \frac{1}{\gamma_{a_-}} \left[ \gamma + \frac{V_0^2}{\mathit{\Gamma}/4} \abs{ 1 - \frac{\alpha^2 g_0^2}{\mathit{\Gamma}\tilde{\kappa}/2}}^2 \right] T_\textrm{bulk} \nonumber \\
	&=&  \frac{1}{\gamma_{a_-}} \left[ \gamma + \frac{4V_0^2}{\mathit{\Gamma}} \frac{(\omega - \mathit{\Delta})^2 + \kappa^2/4}{(\omega - \mathit{\Delta})^2 + (\kappa/2 + 2\alpha^2 g_0^2/\mathit{\Gamma})^2} \right] T_\textrm{bulk}
	\eea
	When  near resonance, $\omega \approx \mathit{\Delta}$, we have:
	\bea
	T_{a_-}^{\text{eff}}& =&  \frac{1}{\gamma_{a_-}} \left[ \gamma + \frac{4V_0^2}{\mathit{\Gamma}} \frac{\kappa^2/4}{(\kappa/2 + 2\alpha^2 g_0^2/\mathit{\Gamma})^2}  \right] T_\textrm{bulk} \nonumber \\
	&=&\frac{1}{\gamma_{a_-}} \left[ \gamma + \frac{4V_0^2}{\mathit{\Gamma}} \frac{1}{(1 + \mathcal{C}_\alpha)^2}   \right] T_\textrm{bulk} \label{eq:20}
	\eea
	It reveals that the second term in equation (\ref{eq:20}) decreases with increasing $\mathcal{C}_\alpha$. This fact indicates the effective temperature of the ccw $a_{-}$ mode reduces with increase of the cw pump laser ($\propto |\alpha|^2$). 
	We can derive the effective temperature of the $a_+$ mode in the same manner. 
	\be
	T_{a_+}^{\text{eff}} =  \frac{1}{\gamma_{a_+}} \left[ \gamma + \frac{4V_0^2}{\mathit{\Gamma}}  \frac{1}{(1 + \mathcal{C}_\beta)^2}  \right] T_\textrm{bulk}. 
	\ee
	In the experiment shown in the main paper, the ccw probe $\beta$ is much smaller compared to the cw pump $\alpha$, thus this effective temperature $T_{a_+}^{\text{eff}}$ change is not significant for the $a_+$ mode.

	\vspace{24pt}
	\noindent
	\textbf{SN2.4 Analysis of the direct coupling model}
	\vspace{12pt}
	
	For the direct coupling model, i.e. disorder only coupling $a_-$ and $a_+$ via $V_1$ and no additional quasi-modes, we can find striking differences from the observations (See Supplementary Fig.~\ref{fig:s2}). 
	After adiabatic elimination of $c_\pm$, we have Heisenberg-Langevin equations for $V_0 = 0$ of
	\begin{align}
	\dot{a}_- &= -[i \omega_m + (\gamma + \mathit{\Gamma}_\beta)/2] a_- + \sqrt{\gamma} a_{-,in} - i V_1 a_+ \\
	\dot{a}_+ &= -[i \omega_m + (\gamma + \mathit{\Gamma}_\alpha)/2] a_+ + \sqrt{\gamma} a_{+,in} - i V_1 a_- 
	\end{align}
	with $\mathit{\Gamma}_\alpha \equiv 4 h_0^2 |\alpha|^2/\kappa$ the optically-induced damping. We see that the normal modes of these equations have resonance conditions corresponding to two poles:
	\begin{equation}
	\omega_{\pm} = \omega - i \mathit{\Gamma}_\Sigma/2 \pm \sqrt{V_1^2 - \delta \mathit{\Gamma}^2/4}
	\end{equation}
	where $\mathit{\Gamma}_{\Sigma} = \gamma + \frac{\mathit{\Gamma}_\beta + \mathit{\Gamma}_\alpha}{2}$ is the average damping and $\delta \mathit{\Gamma} = |\mathit{\Gamma}_\alpha - \mathit{\Gamma}_\beta|$ is the difference in damping. 
	Thus at zero power the two poles are split on the real axis by $\pm V_1$, leading to mode splitting which is not observed in the experiment. Furthermore, as $\delta \mathit{\Gamma}$ increases to be larger than $V_1$, the damping rates start to differ, whereas in the experiment the damping is different for all optical powers. Finally, at high $\delta \mathit{\Gamma}$, the imaginary (damping) part of the pole is still always $\geq \gamma$, the value of the damping at zero optical power in this model, counter to the observed behavior in the experiment. Regarding the temperature of the $a_-$ mode, working in the large $\delta \mathit{\Gamma}$ limit, we do see some cooling of $a_-$ at intermediate powers, as predicted in Supplementary Ref.~\cite{xunnongCooling16}.

	\begin{figure}[!hp]
		\includegraphics[width=0.8\textwidth]{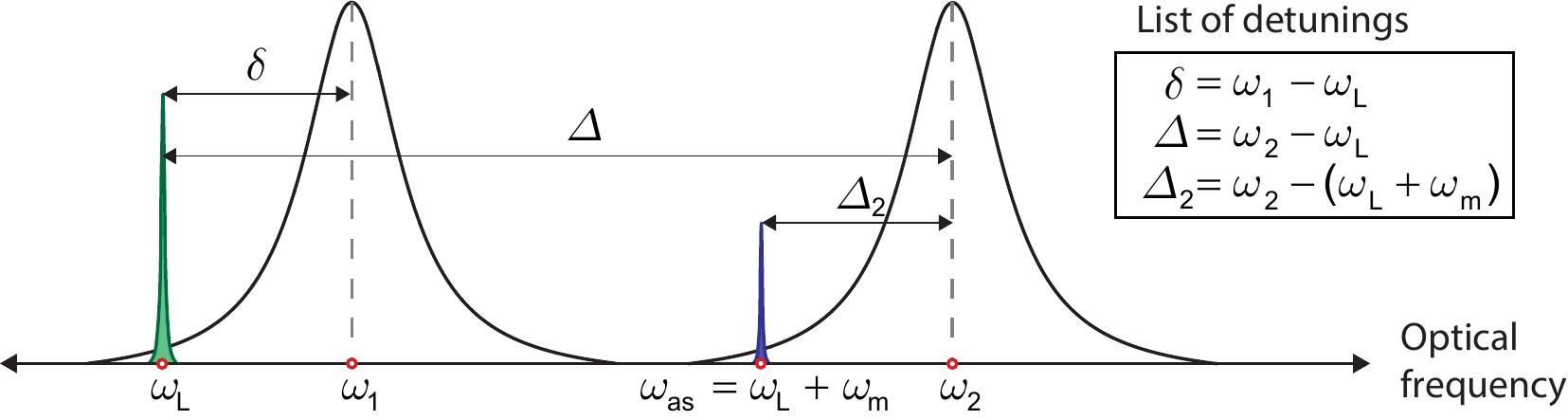}
		\centering
		\caption{Optical frequency relationship of the triplet resonant system for the experiment. The pump laser is at frequency $\omega_L$, and the anti-Stokes scattered light appears at frequency $\omega_{as} = \omega_L + \omega_m$ via Brillouin scattering. The two frequency-adjacent optical modes are at $\omega_1$ and $\omega_2$, and the corresponding signal detunings are defined as $\delta = \omega_1 - \omega_L$, $\mathit{\Delta} = \omega_2 - \omega_L$ and $\mathit{\Delta}_2 = \omega_2 - \omega_{as}$. }
		\label{fig:s3}
	\end{figure}

	\begin{figure}[!hp]
		\includegraphics[width=0.8\textwidth]{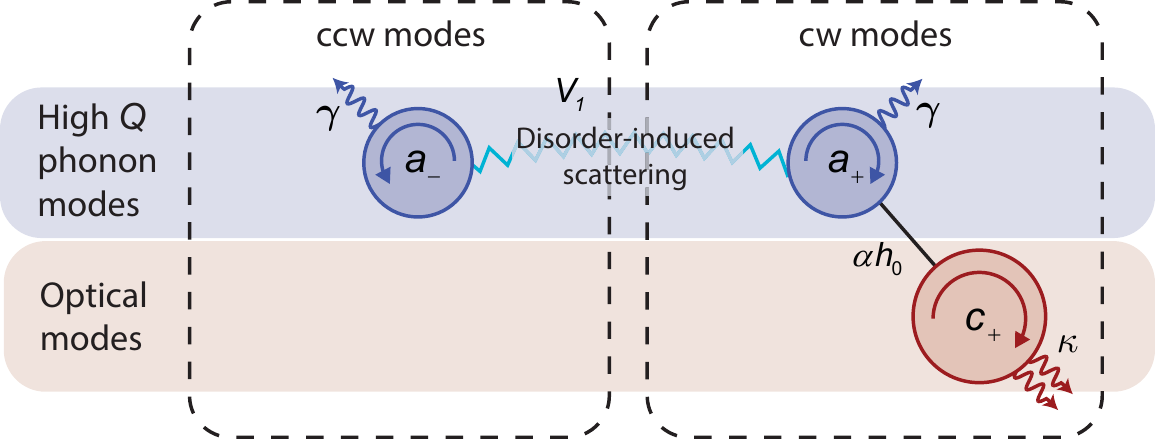}
		\centering
		\caption{The `direct coupling model' consists of the optomechanical coupling of the $c_{+}$ optical mode to the $a_{+}$ high-$Q$ mode, and direct disorder-induced coupling between the $a_{\pm}$ modes. $\alpha h_0$ is the light-enhanced optomechanical coupling strength and $V_1$ is the direct coupling rate.}
		\label{fig:s2}
	\end{figure}

	\FloatBarrier

	\FloatBarrier
	
	\newpage
	
	\begin{center}
		\textbf{\large Supplementary Note 3. Experimental details} 
	\end{center}
	
	In Supplementary Fig.~\ref{fig:s4}, we illustrate the detailed experimental setup used to measure the spectra of the cw (ccw) high-$Q$ phonon modes $a_{\pm}$. The experiment is performed using a silica microsphere resonator optical $Q>10^8$ that is evanescently coupled to a tapered fiber waveguide. The scattered light that results from the opto-acoustic coupling is sent to the photodetector through the same waveguide that also carries the pump laser. A tunable External Cavity Diode Laser (ECDL) spanning 1520 - 1570 nm drives light into the waveguide. A 90/10 fiber optic splitter separates this source into the cw pump and the ccw probe laser. 
	In the cw direction, the pump laser is amplified by an Erbium-Doped Fiber Amplifier (EDFA). Thus, the EDFA affects the cw pump laser only, not the ccw probe laser.

	In order to measure anti-Stokes light detuning from the optical mode, $\mathit{\Delta_2} =  \omega_2 - \omega_{AS}$, we employ a Brillouin Scattering Induced Transparency measurement~\cite{Kim2015}.
	Here $\omega_{2}$ is the resonant frequency of the anti-Stokes optical mode, and $\omega_{AS}$ is the frequency of anti-Stokes scattered field via Brillouin scattering as illustrated in Supplementary Fig.~\ref{fig:s3}.
	An electro-optic modulator (EOM) is used to generate the required probe sidebands relative to the cw pump laser. The upper sideband is used to probe the acousto-optic interference within anti-Stokes mode. 

	$1~\%$ of the signal after the EOM output is used as a reference to a network analyzer (NA) to measure the transfer function for this optical probe.
	The remaining light passes through a fiber polarization controller (FPC) to maximize coupling between the taper and the resonator. 
	Circulators are employed for performing analysis of the cw and ccw scattered light. An oscilloscope (OSC) and a real-time spectrum analyzer (RSA) are used to measure these optical signals on a photodetector.

	\begin{figure}[hp!]
		\includegraphics[width=0.8\textwidth]{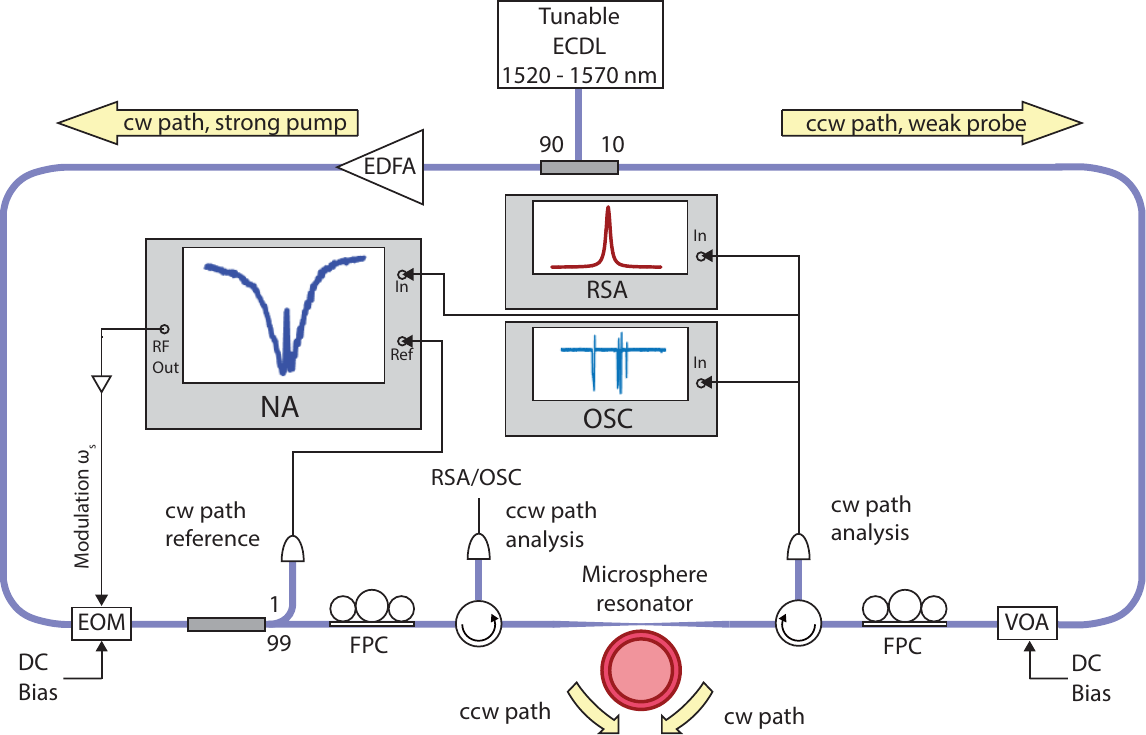}
		\centering
		\caption{Detailed experimental setup. The blue lines indicate the optical paths (fiber), while the black narrow lines indicate the electrical signal paths. A fiber-coupled tunable external cavity diode laser (ECDL) provides light through a 90/10 optical coupler that splits the light into cw and ccw directions. The cw pump power is controlled by an Erbium-doped fiber amplifier (EDFA). An electro optic modulator (EOM) is employed for probing detuning of anti-Stokes light from its optical mode. Scattered light is captured at a photodetector and is sent to an oscilloscope (OSC) and a real-time spectrum analyzer (RSA) for analysis.}
		\label{fig:s4}
	\end{figure}

	\FloatBarrier

	\vspace{24pt}
	\begin{center}
		\textbf{\large Supplementary Note 4. \\Measuring phonon mode spectra using a photodetector}
	\end{center}
	
	To experimentally confirm the reduction of intrinsic damping of the high-$Q$ phonon mode, we have to understand the measurement of the output spectrum at the photodetector.
	Using the equations (\ref{eq:1}, \ref{eq:18}), we can rewrite the Heisenberg-Langevin equations for the $a_-$ phonon mode in frequency domain with the non-depleted pump approximation:
	\begin{subequations}
		\begin{align}
		-i\omega c_{p, \sigma} &= -i\delta c_{p, \sigma}  -\frac{\kappa_{\mr{p}}}{2}  +\sqrt{\kappa_{\mr{ex}}}S,\\ 
		-i\omega c_- &= -i\Delta c_-  -\frac{\kappa}{2} c_- -i \beta h_0 a_- + \sqrt{\kappa_{\mr{ex}}}c_-^{\mr{in}},\\ 
		-i\omega a_- &= -i\omega_m a_-  - \frac{\gamma}{2} a_- - i \beta h_0 c_- + \sqrt{\gamma}a_-^{\mr{eff}},\\
		\textrm{where~} \notag 
		\sqrt{\gamma} a_-^{\text{eff}} &= \sqrt{\gamma} a_-^{\text{in}}  - \frac{iV_0}{\sqrt{\mathit{\Gamma}}/2} \left( 1 - \frac{\alpha^2 g_0^2}{\mathit{\Gamma}\tilde{\kappa}/2} \right) b_+^{\text{in}}. 
		\end{align}
		\label{eq:22}
	\end{subequations}
	Here the stationary ergodic noise forces $S$, $c_-^{\mr{in}}$ and $a_-^{\mr{eff}}$ are the quantum Langevin noise of the $c_{p, \sigma}$, $c_-$ and $a_-$ modes, respectively. The quantum correlation functions of these noise forces are given by:
	\begin{align*}
	\langle S^{\dagger}(t) S(t') \rangle &= n_L (t' - t), \\	
	\langle S(t) S^{\dagger}(t') \rangle &= (n_L+1) (t' - t), \\	
	\langle c_-^{\mr{in}~\dagger}(t) c_-^{\mr{in}}(t') \rangle &= 0, \\	
	\langle c_-^{\mr{in}}(t) c_-^{\mr{in} ~\dagger }(t') \rangle &= \delta( t - t'), \\
	\langle a_-^{\mr{eff}~\dagger}(t) a_-^{\text{eff}}(t') \rangle &= \bar{n}\delta( t - t'), \\
	\langle a_-^{\mr{eff}}(t) a_-^{\text{eff} ~\dagger}(t') \rangle &= (\bar{n} + 1)\delta( t - t')
	\end{align*}	
	where $n_L$ is the photon occupation number from the pump laser and $\bar{n}$ is the effective occupation number of phonons. We then obtain the noise spectrum of the $a_-$ mode as follows:
	\begin{align}
	a_- (\omega) = \frac{\sqrt{\gamma} a_-^{\mr{eff}} (\omega)}{\gamma_\mr{e}/2 +i(\omega_m -\omega)} - \frac{i \beta h_0 \sqrt{\kappa_{\mr{ex}}}c_-^{\mr{in}} }{\left[ \gamma_\mr{e}/2 +i(\omega_m -\omega)\right] \left[ \kappa/2 - i(\omega - \mathit{\Delta}) \right]}
	\end{align}
	where $\omega_m$ is the effective mechanical frequency including the optical spring effect and $\gamma_\mr{e} =\nobreak \gamma + \gamma_{\rm{opt}}$ is the effective mechanical damping rate including the optomechanical damping rate~$\gamma_{\rm{opt}}$.
	We can then derive the output spectrum measured at the downstream photodetector after the resonator. Using the input-output theory~\cite{gardiner2004quantum}, we obtain the expression for the output field in the optical waveguide. 
	
	\begin{align}
	\begin{split}
	S_{\mr{out}}(\omega) = &~S_{in}(\omega) -\sqrt{\kappa_{\mr{ex}}} c(\omega) \\
	=&~S(\omega) \left[ 1- \frac{2\kappa_{\mr{ex}}}{\kappa_{\mr{p}} - 2 i (\omega - \delta) } \right]\\ 
	&+ c_-^{\mr{in}} (\omega) \left[ 1 - \frac{2 \kappa_{\mr{ex}}}{\kappa - 2 i (\omega -\mathit{\Delta} )} 
	+ \frac{|\beta|^2 |h_0|^2 \kappa_{\mr{ex}}}{\left[ \gamma_\mr{e}/2 +i(\omega_m -\omega)\right] \left[ \kappa/2 - i(\omega - \mathit{\Delta}) \right]^2}\right]\ \\
	&+ a_-^{\text{eff}}(\omega) \frac{i\beta h_0 \sqrt{\gamma} \sqrt{\kappa_{\mr{ex}}}}{\left[ \gamma_\mr{e}/2 +i(\omega_m -\omega)\right] \left[ \kappa/2 - i(\omega - \mathit{\Delta}) \right]}\\
	=&~s_1(\omega) S(\omega) + s_2(\omega) c_-^{\mr{in}}(\omega) +s_b(\omega)a_-^{\mr{eff}}(\omega)
	\end{split}
	\end{align}
	where we are introducing the scattering matrix elements $s_1(\omega)$, $s_2(\omega)$ and $s_b(\omega)$ defined in~\cite{Safavi-NaeiniPainter2011}. The output spectrum at the photodetector is related to the spectrum of the normalized photocurrent $S_{II} (\omega')\delta(\omega - \omega')  = \langle I(\omega)^\dagger I(\omega')\rangle$ where $I(\omega) = S_{\mr{out}}(\omega) + S_{\mr{out}}^\dagger (\omega')$. Thus, $S_{II}(\omega)$ is:
	\begin{align}
	S_{II}(\omega) &= \left( |s_1(\omega)|^2 + |s_2(\omega)|^2 + |s_b(\omega)|^2\right) +2n_L |s_1(\omega)|^2+ 2 \bar{n} |s_b(\omega)|^2
	\end{align}
	The phonon noise spectrum $S_{a_{-}}(\omega) = \frac{\bar{n}\,\gamma}{(\gamma_e/2 )^2 + (\omega_m -\omega)^2}$ is included in the above expression through the scattering element $2\bar{n} |s_b(\omega)|^2$, since $2\bar{n} |s_b(\omega)|^2  = \frac{2\gamma_{\mr{opt}} \kappa_{\mr{ex}}}{\kappa} \frac{\bar{n}\,\gamma}{(\gamma_e/2 )^2 + (\omega_m -\omega)^2}$.
	The remainder of the equation constitutes the noise floor $N$, which is a function of $n_L$. The resulting photocurrent spectrum is given by:
	\begin{align}
	S_{II}(\omega) &= N + \frac{2\gamma_{\mr{opt}} \kappa_{\mr{ex}}}{\kappa} S_{a_{-}}(\omega) 
	\end{align}
	Thus, the measured RF output spectrum at the photodetector (ignoring noise floor N) is proportional to the spectrum of the high-$Q$ phonon mode $a_-$, scaled by the optomechanical damping rate $\gamma_{\rm{opt}} = \frac{4|\beta|^2|h_0|^2 }{\kappa}$ when $\omega \approx \omega_m$. Fixing the ccw probe power while measuring the spectrum of the ccw high-$Q$ phonon mode $a_-$ ensures that the magnitude scaling of the output spectrum is not affected by the ccw probe power. Thus, the spectrum obtained in $S_{II}(\omega)$ is directly representative of the phonon population and temperature of the mode.

	\newpage
	\renewcommand{\section}[2]{}%
	
	\begin{center}
		\textbf{\large Supplementary References}
	\end{center}

	\def\url#1{}

\end{document}